\documentclass[sigconf]{acmart}

\AtBeginDocument{%
  \providecommand\BibTeX{{%
    \normalfont B\kern-0.5em{\scshape i\kern-0.25em b}\kern-0.8em\TeX}}}

\usepackage{graphicx,subfigure}

\makeatletter
\def\@copyrightspace{\relax}
\makeatother

\begin{document}


\newcommand{\rg}[1]{\textcolor{orange}{\bf\small [#1 --RG]}}
\newcommand{\sans}[1]{\textcolor{blue}{\bf\small [#1 -SS]}}
\newcommand{\nt}[1]{\textcolor{red}{\bf\small [#1 --NT]}}

\title{User Persona Identification and New Service Adaptation Recommendation }

\author{Narges Tabari}
\email{nargesam@amazon.com}
\affiliation{%
  \institution{AWS AI Labs}
  \city{Santa Clara}
  \state{CA}
  \country{USA}
  }
  
\author{Sandesh Swamy}
\email{sanswamy@amazon.com}

\affiliation{%
  \institution{AWS AI Labs}
  \city{Seattle}
  \state{WA}
  \country{USA}
}
  
\author{Rashmi Gangadharaiah}
\email{ rgangad@amazon.com}

\affiliation{%
  \institution{AWS AI Labs}
  \city{Santa Clara}
  \state{CA}
  \country{USA}
}


\renewcommand{\shortauthors}{Tabari, et al.}

\begin{abstract}
Providing a personalized user experience on information dense webpages helps users in reaching their end-goals sooner. We explore an automated approach to identifying user personas by leveraging high dimensional trajectory information from user sessions on webpages. While neural collaborative filtering (NCF) approaches pay little attention to token semantics, our method introduces SessionBERT, a Transformer-backed language model trained from scratch on the masked language modeling (mlm) objective for user trajectories (pages, metadata, billing in a session) aiming to capture semantics within them. Our results show that representations learned through SessionBERT are able to consistently outperform a BERT-base model providing a 3\% and 1\% relative improvement in F1-score for predicting page links and next services. We leverage SessionBERT and extend it to provide recommendations (top-5) for the next most-relevant services that a user would be likely to use. We achieve a HIT@5 of 58\% from our recommendation model.

\end{abstract}





\maketitle


\section{Introduction}


Automatically identifying and determining a user’s preferences on a webpage is an important step towards building personalized systems that satiate a user’s business needs. 
A user’s journey through a website can be emitted in the form of session-based information consisting of atomic semantic units such as the title of the page clicked, the text content of a button that was clicked, text within help articles, location information, billing information, and so on. Leveraging such semantic information through the course of a user’s journey can give us valuable insights into their preferences. 

Depending on the complexity of tasks, user sessions will span anywhere from a few minutes to several hours filled with actions necessary to reach a desired goal. User facing interfaces tend to consist of dynamically generated widgets such as \textit{Recently Visited}, \textit{System Health}, and \textit{Explore new services}. Recommending the right widget, along with personalized content will help users to explore the right services and functionalities to arrive at their end task much sooner. As an added bonus, this allows users to adopt previously unknown capabilities, thereby ensuring a smoother user experience. As a first step towards achieving this, we explored the idea of assigning Personas\footnote{Using user survey data, seven different personas were defined for our users. Each persona was assigned a set of tasks as their main activity. These seven personas share tasks between each other, and have a distinct set of tasks that is only assigned to them. For example,  \textit{Persona 1} has \textit{monitoring} as one of its tasks, compared to \textit{Persona 2} that contains \textit{cost management} as one of their core tasks, however, both of them share \textit{exploration of new services} as their tasks as well. For anonymity, we name these personas Persona 1 to 7. } to our users through \textit{Persona Identification} where we detect \textit{existing} and \textit{emerging} user personas on the website. We leverage techniques from Natural Language Processing (NLP) and language modeling (LM) \cite{StanfordMM} to incorporate semantics and context into our model in the form of user journeys consisting of page titles, location data, and billing information. LMs in general can give personalized recommendations corresponding to historical behaviors. Obtaining a persona helps understand a user’s intention, generate a user profile for each of them, and be able to provide proactive personalized experiences and recommendations. Our recommendation engine recommends a personalized list of top 5 services previously unexplored by users.

The goal of persona identification is to segment users based on their activities on the user interface.  
To assign users to different segments, we need to learn representations that can succinctly summarize the user's behavior. 
Due to lack of labeled data, we approach persona identification in an unsupervised setting consisting of 3 stages. First, we generate activity-based, semantic session representations using the click stream data.
Motivated by language models leveraged in NLP applications \cite{10.1145/3477495.3531810}, we introduce \textbf{SessionBERT}, a pre-trained Transformer-backed language model trained from scratch that takes advantage of semantic information in text such as user interface clicks, geo-location, billing, and different activity features to understand a user behavior. We train the model with the same Masked Language Modeling optimization, but we mask a user's session information in our approach. The resulting model is further fine-tuned in a multi-task setting in order to evaluate the quality of learned representations along with their prediction power on the last service and last page prediction in each session sequence. Second, we group users based on the learnt session embeddings from SessionBERT. 
Finally, using high-level descriptions of personas defined through user surveys, we map the clustered groups to personas. 

For new service recommendation, we used the SessionBERT model and fine-tuned it on service and page prediction tasks. The goal of this recommendation is for users to adapt a new service that they have not used before. We show that SessionBERT fine-tuned on service prediction and ranking recommendations based on the number of times it is recommended is able to provide contextual recommendation to users. We deploy all our models using Sagemaker pipelines and use p3.2xlarge instances to host. We use dynamic widgets placed on user homepages to surface service recommendations.

Our main contributions are:
\begin{itemize}
\item SessionBERT: We introduce SessionBERT, a pre-trained Transformer based language model, trained from scratch on user session activities on a webpage. The proposed approach provides a substantial gain of 29\% relative for service prediction and 32\% relative for page prediction (for sequence length of 64) and 1\% relative for service and 3\% relative for page prediction in F1-measure (for sequence length of 512) over the BERT-base model.  
\item User Persona Identification: We present a method to create user personas using the representations from SessionBERT through three phases: building session representations, user segmentation, and  persona mapping. 
\item New Service Recommendation: We show that SessionBERT fine-tuned on service prediction and ranked based on most services recommended has 93.1\% accuracy and can provide contextualized recommendation tailored to user history and their persona.
\end{itemize}


\section{Related Work}
\label{sec:work}
One of the main challenges in personalization is to generate appropriate content for each user in order to assist them and gain trust \cite{unknown}. Existing personalized language models can be classified into two kinds: explicit and implicit personalization \cite{article}. Explicit personalization models introduce profiles and personal intersests such as gender, and expertise into language models to generate personalized responses. On the other hand, implicit personalization models extract persona profiles using user’s activities and all their historical actions which is our focus.

Research around developing implicit personalization models and systems on user-specific textual and session content is under-explored. A natural approach used in recent years has been to train a language model for each user, based on their historical activities\cite{oba-etal-2019-modeling}, \cite{10.1145/3477495.3531810}. For example, \cite{oba-etal-2019-modeling} proposed a personalized review generation model by considering different reviewers and their style and choice of vocabulary. Or \cite{10.1007/s11280-018-0598-6} first pre-trained a general response generation model on large-scale conversational data, and then personalized the responses to user’s preferences by tuning the model on their personal conversation data.

User modeling, especially with goal of recommendation is a fundamental approach for implicit personalization tasks such as personalized news \cite{10.1145/3097983.3098108}, or item recommendation \cite{10.1145/3357384.3357895}. Most of the existing user modeling techniques rely on labeled data \cite{10.1145/3397271.3401156} for training accurate models. Motivated by pre-trained language models in NLP, these applications pre-train self-supervised models on user behavior data and fine-tune its parameters on the desired downstream tasks such as item recommendation \cite{10.1145/3477495.3531810}, \cite{ DBLP:journals/corr/abs-2012-06678}. Most of implicit personalization models, however, either have been focused on specific application such as dialog systems, or in a more general way, they usually aggregate all user behavior data in one single data sample \cite{10.1145/3477495.3531810}. Most recently, there are some groundwork attempts that
uses Large Language Models (LLMs) for solving item recommendation tasks \cite{wang2023generative, wang2023zeroshot}. However, these models are very costly to train and are not suitable for low latency requirements of our work. Our work leverages LMs to generate user representations, provide user segmentations, and provide recommendations.

\section{Data}
\label{sec:data}

\begin{figure*}[!htb]
\begin{centering}
\includegraphics[scale=0.3]{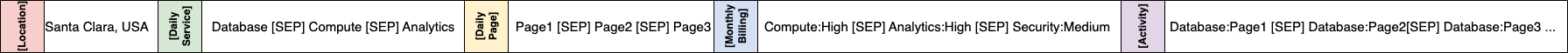}
\caption{Sample data point used in training of SessionBERT.}
\label{fig:sample_data}
\end{centering}
\end{figure*}


When users interact with 
a web interface, they eventually look to navigate to pages that help them achieve their goal (in our case, service pages). Predicting and recommending next pages to visit helps reduce time spent on the website while providing the benefit of introducing users to new services. A user's session history helps generate representations which capture personalized features such as a user's past transactions, pages visited, billing information (\emph{Low}, \emph{Medium}, \emph{High}, and \emph{Very High}), and geographical locations. For our experiments, we gathered ~9MM session trajectory information. We split the dataset into \textit{5.9MM:1.9MM:1.1MM} sessions for train:validation:test sets respectively. 

These sessions include services, pages, and subpages that each user navigated to reach their goal. Our model for learning user's representations is trained by flattening their session history into a sequence of \emph{service;page} combinations.
Each activity is separated by \textit{[SEP]} token. Additionally, we add user billing information, and geo-location as context to our session sequence. Figure \ref{fig:sample_data} provides a sample for the data that acts as input to our models. 

\section{Model}
\label{sec:model}

For all experiments  we use 4 NVIDIA Tesla V100 GPUs. All models run for 3 epochs with learning rate of $2\mathrm{e}{-5}$. For pre-training we masked 15\% of tokens in the user trajectories. 

\subsection{SessionBERT}

Using data from Section \ref{sec:data}, we trained a tokenizer (vocabulary size of 30k) on user's session information which is then used to train a BERT model \cite{devlin-etal-2019-bert} using Masked Language Modeling (MLM) loss but optimizing the model to be able to predict missing web/service pages instead. 
Additional tokens were added when creating the vocabulary which help separate different sections of a training instance such as \emph{[activity]} for sequence of service;page combination, \emph{[daily\_page\_token]} for daily visited pages, \emph{[location\_token]} for country and city section, \emph{[daily\_service\_token]} for daily visited service pages, \emph{[daily\_billed\_token]} for daily most billed services, and \emph{[monthly\_billed\_token]} for monthly most billed services.  

\subsection{User Segmentation}
\label{sec:cust_segmentation}

\begin{figure}[h]
  \centering
  \begin{minipage}[b]{0.2\textwidth}
    \includegraphics[width=\textwidth]{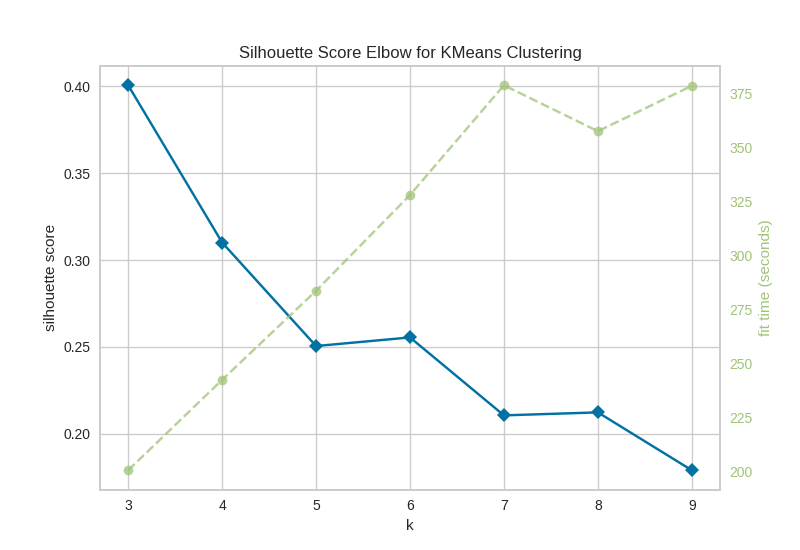}
    
  \end{minipage}
  \hfill
  \begin{minipage}[b]{0.2\textwidth}
    \includegraphics[width=\textwidth]{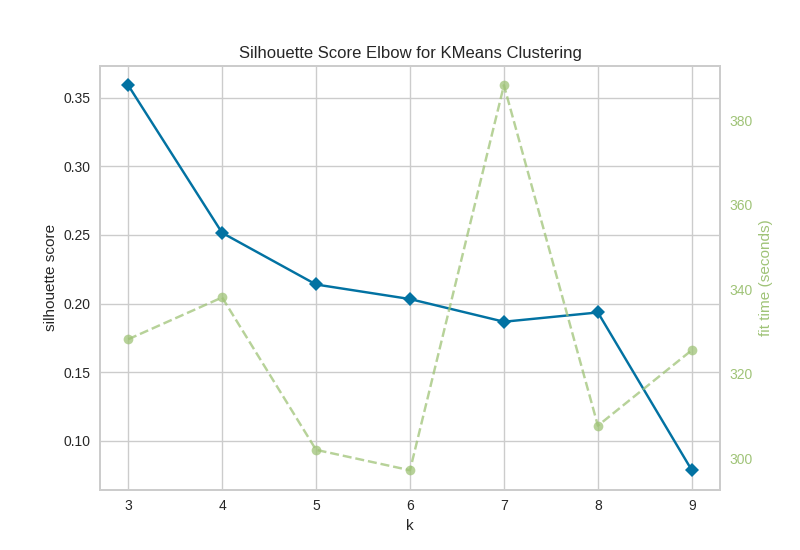}
    
  \end{minipage}
  \caption{Elbow method results for SessionBERT (left) and Bert-base (right) embedding for number of clusters K = 3 to 9. The Blue line is Silhouette score and green is run-time.}
    \label{fig:SessionBERT}

\end{figure}

Using the embeddings obtained from SessionBERT, we segmented session representations using K-means \cite{Hartigan1979} clustering using cosine similarity 
as our distance metric. To determine the number of clusters and to additionally evaluate SessionBERT embeddings, we experimented with different values of $k$ ($k=3-9$) for both BERT-Base \cite{devlin-etal-2019-bert}, and SessionBERT and calculated the silhouette score \cite{Peter} for each run. 

As shown in Figure \ref{fig:SessionBERT}, the silhouette score for SessionBERT outperforms BERT-Base for all values of $k$. Additionally, by looking at the plot for SessionBERT we can observe that $k=4$ is an appropriate number of clusters for segments in our data.





    
    

\subsection{Mapping clusters to pre-defined personas}

On obtaining segments ($k=4$) as described in Section \ref{sec:cust_segmentation}, we perform a mapping of each segment to one or more pre-defined personas ($7$ in number). 
First, we manually mapped the set of top-200 activities performed by all users on the webpage to pre-defined persona tasks. An \emph{activity} is defined as a combination of [service:page] states that a user visits through the course of different sessions. Eg: the activities \textit{"service1;page10"} and \textit{"service6;page14"} is mapped to the task \textit{Setup Environment}. We then identified the top ten activities performed in each cluster. The earlier mapping helped determine the personas captured by each of the clusters.

\paragraph{Embedding of personas and clusters given tasks} To map pre-defined personas to clusters, we generated a one-hot encoding for each persona as follows: we 
set the dimension $d$ of the embedding to be the total number of identified tasks. Then for each persona we edit the encoding to $1$ if a task exists in a persona definition. A similar encoding was formulated for each cluster, but for each task's index, we calculated probability of each persona's task being used in each cluster using the following formula:

\[p(task_i|cluster_k)=\frac{num(task_i, cluster_k)}{\sum_j num(task_i, cluster_j)}\]

Using these probabilities, for each \(cluster_{k}\), an embedding was created that if a persona task exists in that cluster, then the probability of that task was assigned at the index. 
The probability distribution of clusters being assigned to a specific persona was calculated as follows:

\[(cluster_k , persona_i) = cluster_k\quad  . \quad  persona_i\]

and the persona with maximum probability was picked for the cluster as its persona tag.

\[p(cluster_k|persona_i)=\max(\sum_k(cluster_k, persona_i))\\ 
\forall i: 0,1,2,3\]

The final mapping of personas and clusters are presented in Table \ref{tab:persona mapping}. For every new user at inference time, the distance between their session's embeddings (as computed by SessionBERT) and the cluster centroids was calculated and the closest cluster (persona) was assigned. 

As observed in Table \ref{tab:persona mapping}, for cluster 2 there were two different personas that 
could be assigned. We merged these two personas and assigned both to a single cluster. Additionally, persona 6 and 7, had extremely low probability of being assigned to any cluster. Therefore, it was evident that there was not enough signal in session activity for them.

\begin{table}[h]{}
		\centering
	\caption{Personas mapped to each cluster.}
            \begin{tabular}{ccl}
            \hline
            \textbf{Cluster} & \textbf{Persona} \\ \hline
            0 &  Persona 1 \\ \hline
            1 & Persona 3 \\ \hline
            2 & Persona 2 and Persona 4  \\ \hline
            3 &  Persona 5  \\\hline
            \end{tabular}
		
		\label{tab:persona mapping}
	\end{table}
\subsection{Human Evaluation of Personas}

To evaluate the quality of the personas assigned to users based on our models, four annotators were given 179 sessions that were mapped with the personas inferred. Annotators were tasked to pick an annotation for each from the following labels: \textit{Match}, \textit{Not\_matched}, and \textit{Unclear}. We found that across all persona clusters, in 74.25\% of the instances, all annotators agreed that the persona is appropriately assigned to the user session data. The details about human evaluation is included in Appendix \ref{persona_human}.

\subsection{New Service Recommendation} 
We leverage representations learnt from SessionBERT to provide next service recommendations to users. To evaluate our model's ability in providing recommendations, we used the last activity (service and page) of each session as the ground truth labels to be predicted for each session sequence. As a baseline, we fine-tuned the BERT-Base model \cite{devlin-etal-2019-bert} on service and page prediction task using a multi-task setting. 

Our results in Table \ref{tab:multitask}, show that SessionBERT fine-tuned on next service and page prediction outperforms Bert-base on both tasks. The label space for recommending new pages ($2288$) is much larger than that of services ($190$) which explains the difference in performance.
To proactively recommend new services to a user, we store their last eight sessions on a given day. (We determined that 8 sessions had to be used as context for recommendation via a human evaluation study which can be found in Appendix \ref{eightsession}).

During inference, the top-5 service recommendations were calculated 
for each user from our SessionBERT model and if a service was already used by a user, it was filtered out. Finally, the top 5 new service recommendations was calculated based on the number of times they were recommended to them in our system. We picked \textit{number of occurrence of a new service} as our ranking algorithm based on a human evaluation study explained in Appendix \ref{ranking} .

\begin{table}
  \caption{Multitask learning comparison on SessionBERT and BERT}
  \label{tab:multitask}
  \begin{tabular}{ccccl}
  \hline
\multicolumn{1}{c}{ } 
 & \multicolumn{2}{c}{Service prediction head} 
  &  \multicolumn{2}{c}{Page prediction head} \\
    \toprule
    Model & F1-measure & Accuracy & F1-measure & Accuracy \\
    \midrule
    \ BERT-Base & 0.924 &	0.923 &	0.730 &	0.720\\
    SessionBERT & \textbf{0.932} & \textbf{0.931} & \textbf{0.753} & \textbf{0.730}\\

  \bottomrule
\end{tabular}
\end{table}



\section{Ablation Studies}
\label{sec:results}

\textbf{Sequence Length:} We varied the length of session sequences provided as context to our models. 
As longer spans of information is passed to SessionBERT and BERT-Base, 
we found that the model performance also improved. Regardless of the length picked, SessionBERT consistently outperforms BERT-Base. It takes 10-fold decrease in fine-tuning time for sequence length of 64 compared to 512, and so therefore SessionBERT is especially helpful when limited computational resources exists as 
the results for SessionBERT compared to BERT-Base for length 64 is more pronounced. SessionBERT provides an improvement of 29\% relative for service prediction and 32\% relative on page prediction on F1-measure for sequence length of 64. Table \ref{tab:service length} and Table \ref{tab:page length} show all the results for this experiment. 

		
           
		
            

           
	

\begin{table}[h]
		\centering
		
		\centering
	\caption{Service Prediction: Study of session sequence length. Numbers shown here are F1-measure. }
            \begin{tabular}{ccccl}
            \hline
             & \textbf{64} & \textbf{128} & \textbf{512}\\ \hline
            BERT-Base &  	0.63 &	0.90 &	0.920 \\ \hline
            SessionBERT & 0.818 &	0.923 &	0.931 \\ \hline
           
            \end{tabular}
		
		\label{tab:service length}
            
\end{table}

	 \begin{table}[h]
		\centering

	\caption{Page Prediction: Study of session sequence Length. Numbers shown here are F1-measure.}
            \begin{tabular}{ccccl}
            \hline
             & \textbf{64} & \textbf{128} & \textbf{512}\\ \hline
            BERT-Base &  0.368 &	0.628 &	0.72 \\ \hline
            SessionBERT & 0.488 &	0.704 &	0.73 \\ \hline
           
            \end{tabular}
		\label{tab:page length}
	
\end{table}

\textbf{Multitask learning vs separate training}: To measure the impacts of training regimens, we focused on multitask learning setting versus separate fine-tuning for service prediction and page prediction. As shown in Table \ref{tab:multitask vs separate}, we observed that both SessionBERT and BERT-Base benefited from the multitask learning approach on both service and page prediction. 

\begin{table}[h]
  \caption{Multitask learning  vs Separately Training}
  \label{tab:multitask vs separate}
  \begin{tabular}{ccccl}
  \hline
\multicolumn{1}{c}{ } 
 & \multicolumn{2}{c}{Service prediction} 
  &  \multicolumn{2}{c}{Page prediction} \\ 
    \toprule
    Model & F1 & Accuracy & F1 & Accuracy \\
    \midrule
    \ BERT-Base Multitask & 0.924 &	0.923 &	0.730 &	0.720\\
    \ BERT-Base Service & 0.901 & 0.89 & -	 & -\\
    \ BERT-Base Page & - & - &	0.70	& 0.686\\
    SessionBERT Multitask & \textbf{0.932} & \textbf{0.931} & \textbf{0.753} & \textbf{0.730}\\
    SessionBERT Service & 0.893 & 0.892 & - & - \\
    SessionBERT Page & - & - & 0.702 &	0.688\\

  \bottomrule
\end{tabular}
\end{table}

\textbf{SessionBERT recommendation for each Persona category}:

In this experiment, the goal was to check if the recommendations from SessionBERT is personalized towards users based on their personas. For three personas with the most data, we randomly selected 10 users each. These users were assigned a persona by our model with more than 80\% probability. Then for each user in each persona category, we calculated the top-5 recommendations. We found that for the new service recommendations for users in different persona categories, 78\% of the times the services were tailored and specific to the task description of their persona. For example, Persona 1 has ``monitoring" as one of their tasks and all 5 users' recommendations contained new services that help users perform that task.

\textbf{Conversion Rate for SessionBERT recommendations}:
To evaluate the quality of [service;page] recommendations provided by our SessionBERT model, we picked historical data spanning two months for a set of randomly selected users (44k users in total). 
We split the data into ``seen" and ``unseen" partitions based on the number of days (6, 10, and 12 days for ``seen" data). By using the ``seen" partition as historical context, we used our SessionBERT to obtain recommendations for future services. We filter out recommended services that have already been adopted by users previously. Using the ``unseen" partition of our data, we compute the set of services which were eventually adopted by users and compare it with recommendations made using SessionBERT to indicate how often the recommended services are appropriate.

\begin{table}[h]
		\centering
		
		\centering
	\caption{Results of Hit@N experiments for new service adaptation}
            \begin{tabular}{cccl}
            \hline
             & \textbf{Hit@5} & \textbf{Hit@3} \\ \hline
            6 Day split &  	0.54 &	0.42  \\ \hline
            10 Day split & \textbf{0.58} &	0.39  \\ \hline
            12 Day split & 0.55 &	0.32  \\ \hline
           
            \end{tabular}
		
		\label{tab:Hit@N}
            
\end{table}

Our results in Table \ref{tab:Hit@N} show that a split at 10 day ``seen" context provides sufficient contextual clues for our models to perform well on a held-out ``unseen" split. Additionally, we can see that when more data is included in the ``unseen" section of data, it is more likely for the model to predict and recommend more relevant new services. SessionBERT is able to predict new service adaptation with Hit Rate of 58\% for a 10 day ``seen" split for Hit\@5.


\section{Conclusion and Future Work}
\label{sec:conclude}
In this work, we presented a new approach to persona identification and new service and page recommendation for users as part of user-personalized experiences using Transformer-based techniques which leverage Masked Language Modeling (MLM) for recommendations. These personas can be used to recommend services, content, links or pages to users that would help them to achieve their goal faster. Our approach allows accurate mapping of user trajectory data to a set of pre-defined personas in an unsupervised fashion. 
We show that our proposed SessionBert model consistently outperforms BERT-base in service and page prediction tasks. 

\bibliographystyle{ACM-Reference-Format}


\appendix

\section{Human Evaluation on Personas}
\label{persona_human}

We performed human evaluation on result of Persona identification. If the Persona detected by our model was labeled  as \textit{match} by all of our annotators, then we count that as True Positive result of our model, otherwise they were counted as False Negative. Accuracy, Sensitivity, and Specificity were claculated based on TP and FN results. Detailed result of human evaluation per persona is shown in Table \ref{tab:humanevalpersona}: 

\begin{table}[h]
		\centering
		\centering
	\caption{Human Evaluation of session activity on different Personas.  }
            \begin{tabular}{ccccl}
            \hline
             & \textbf{ Accuracy} & \textbf{ Sensitivity} & \textbf{ Specificity}\\ \hline
            Persona 1 &  	0.73 &	0.78 &	0.21 \\ \hline
            Perona 3 & 0.68 &	0.70 &	0.29 \\ \hline
           Persona 2 and 4 &  	0.86 &	0.88 &	0.11 \\ \hline
            Persona 5 & 0.70 &	0.88 &	0.11 \\ \hline
            \end{tabular}
		
		\label{tab:humanevalpersona}
            
\end{table}

\section{Choice of 8 recent sessions}
\label{eightsession}
SessionBERT as a Language Model, can give personalized recommendations corresponding to historical behaviors. At inference time, we provide top-5 service recommendations based on the last session activities of the user. It is important to determine the minimum number of sessions that is appropriate for generating accurate recommendations. For this purpose, we ran experiments on 2 months of data on users with more than 5 sessions per day.


\begin{figure}[!htb]
\begin{centering}
\includegraphics[scale=0.5]{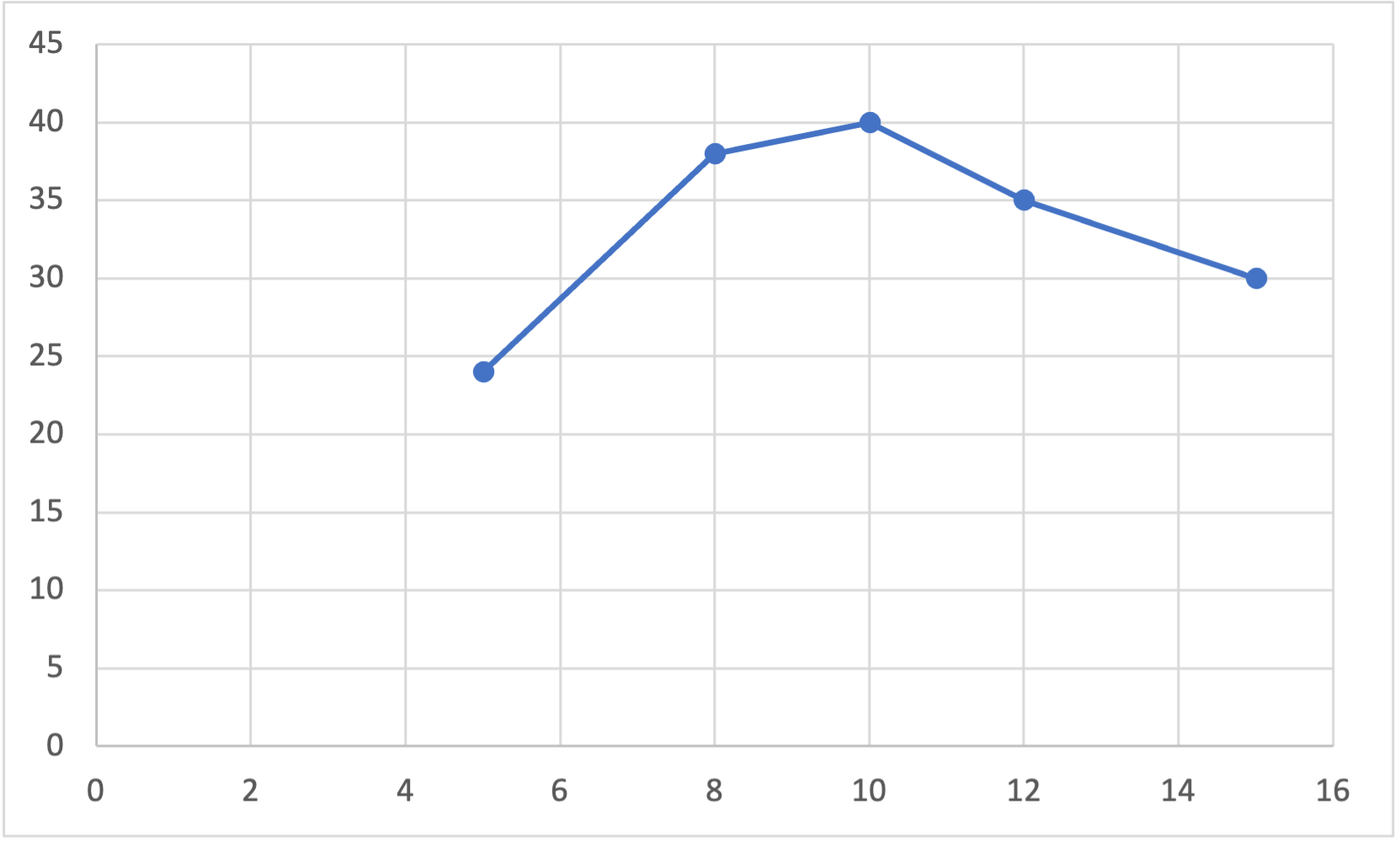}
\caption{This plot shows the number of times each category was picked by annotators as better recommendation.}
\label{historicalsession}

\end{centering}
\end{figure}

For each user, we get their last 3, 5, 8, 10, 12, 15 sessions, and for each session, we inferred top 5 service recommendations, and disregarded the used services for each user. The remaining new services based on last sessions where then manually annotated (not only in terms of quality but also in terms of number of services provided). The result of annotation presented in Figure \ref{historicalsession}. Based on this figure $X=8$ or $X=10$ could both be a good option. Since we want to store the least amount of data as possible, while providing a good quality of recommendation, we chose $X=8$ as our number of historical sessions to continuously store to provide recommendations per user.

\section{Ranking Approaches Human Evaluation}
\label{ranking}
After SessionBERT provides a list of recommendations of new services for each user, the goal is to pick and provide the top 5 new services that is most likely for that user to adapt. For this reason, we performed human evaluation on a randomly selected sample of 50 user session data. For each user, the information of all the services that user has adapted was included. \textit{rec1}, \textit{rec2}, and \textit{rec3} were results of three different types of ranking algorithms. Three annotators were asked to annotate each sample based on the following:  
\begin{itemize}
\item{1:} rec1 is a better recommendation 
\item {2:} rec2 is a better recommendation 
\item {3:} rec3 is a better recommendation
\item {4:} All approaches provides appropriate recommendation
\end{itemize}

\textbf{Ranking approaches:} In \textit{rec1}, the top 5 services were picked based on number of times each service has been recommended to a user. In \textit{rec2}, the top 5 services are picked based on cosine similarity between the name of services to recommend and the name of services the user has used historically. In \textit{rec3} the top-5 services are picked based on cosine similarity between the description of services to recommend and the description of services the account has used historically. 60\% of the times all three annotators agreed that \textit{rec1} was able to provide a better service recommendation to users and 22\% of the time they agreed that all recommendations are equally good.









\end{document}